\documentstyle[emulateapj]{article}
\def\mincir{\ \raise -2.truept\hbox{\rlap{\hbox{$\sim$}}\raise5.truept	
\hbox{$<$}\ }}								
\def\magcir{\ \raise -2.truept\hbox{\rlap{\hbox{$\sim$}}\raise5.truept	
\hbox{$>$}\ }}								

\begin{document}

\title{The Fall of the Quasar Population}

\author{A. Cavaliere$^1$ and V. Vittorini$^{1}$}

\affil{$^1$ Astrofisica, Dip. Fisica 2a Universit\`a, Roma I-00133}

\begin{abstract}

We derive quantitative predictions of the optical and X-ray
luminosity functions for quasars in the redshift range $z \mincir 3$.  
Based on accreting black holes as primary sources for the 
quasar outputs, we investigate how the accretion is controlled by  the
surrounding structures, as these grow hierarchically 
from the formation of the host galaxies to their assemblage 
into poor and eventually into rich groups. 
We argue that for $z < 3$ efficient black hole fueling is triggered by 
the encounters of a gas-rich host with its companions in a group; 
these destabilize the gas and induce accretion, giving rise to the 
following features.
The dispersion of the dynamical parameters in the encounters 
produces luminosity functions with the 
shape of a double power-law. Strong luminosity evolution is produced as 
these encounters deplete the gas supply in the host; 
an additional, milder density evolution 
obtains since the interactions become progressively rarer as 
the groups grow richer but less dense. 
We carry out these arguments to derive a specific model for the evolving 
luminosity functions. From the agreement with the optical and the X-ray 
data, we conclude 
that the evolution of the bright quasars is driven by the development of cosmic 
structures in two ways. 
Earlier than $z \sim 3$ the gas-rich protogalaxies grow by merging, which 
also induces parallel growth of central holes accreting at their full Eddington 
rates. In the later era of group assemblage 
the host encounters with companions drive onto already existing holes 
further but meager accretion; these consume the gas reservoirs in the hosts, 
and cause supply-limited emissions which are intermittent, 
go progressively sub-Eddington and peter out. 
Then other fueling processes occurring in the field come to the foreground; 
we specifically discuss the faint emissions, especially noticeable in X-rays, 
which are expected when hosts in the field 
cannibalize satellite galaxies with their meager 
gas contents.
\end{abstract}

\keywords{Quasars: general -- galaxies: nuclei -- galaxies: interactions -- 
galaxies: clusters: general -- X-rays: galaxies} 

\section{Introduction}

We are motivated by the steep evolution of the bright quasar (QS) population;
out to $z \approx 2.5$ their number exceeds the local value by some $10^{2}$ 
(Schmidt \& Green 1983, Schmidt 1989). 
Around $z\approx 3$ the evolution culminates, and farther out it 
goes into a decline (Osmer 1982, Shaver et al. 1996). 

Our work will be based on two widely shared notions. 
First, the QSs are powered by massive black holes (BHs) accreting gas 
from their host galaxies; thus e.m. outputs  
$$L \sim \eta\, c^2 \Delta m/\Delta t \mincir 10^{48}\; erg\; 
s^{-1} \eqno(1.1)$$
obtain when the baryonic mass $\Delta m$ is accreted over the time $\Delta t$ with 
conversion efficiency up  to $\eta\sim 10^{-1}$ (Rees 1984).
Second, the structures surrounding the BHs develop after the hierarchical 
clustering (see Peebles 1993);  
this envisages the galaxies as baryonic cores within heavier 
halos of cold dark matter (DM). Before $z \approx 2.5$ galactic 
halos are built up through merging of 
smaller units; thereafter the galaxies begin to assemble into small 
groups of dynamical mass $M_G \magcir 5\, 10^{12} \, M_{\odot}$, 
which in turn merge into richer groups and eventually into clusters. 
By such a hierarchical development, the numbers of structures in each 
mass range undergo a fast increase followed by a slow demise. 

On these grounds, the high QS luminosities in the optical, 
IR or X-ray bands may be expected to signal out to 
$z \sim 5$ environmental conditions conducive to accretion. 

In fact, the decline (a rise in cosmic epoch) of the early QSs is widely held 
to be geared to galaxy formation.  Specifically, their environment is 
held to be constituted by protogalactic spheroids  
being built up hierarchically 
through major, chaotic merging events between dense, very gas-rich subunits. 
In these events the gravitational potential is strongly distorted, 
while the baryons are lumped and shocked; 
so the angular momentum $j$ of the gas orbiting at kpc distances 
in the host is efficiently transferred to the heavier DM substructures, 
and the gas can begin an inward course  
toward pc or smaller scales. 
From the plenty of gas so made available, central massive BHs 
collapse and/or accrete rapidly;  
they can accrete at their full {\it self-limiting} rate, 
and easily attain Eddington luminosities $ L_E \approx  10^{47}\, 
M_{BH}/10^9 M_{\odot}$ erg$s^{-1}$. This is indicated by 
eq. (1.1), when the BHs double their mass  
and make $\Delta m \sim M_{BH}$ over short dynamical scales $\Delta t$ 
that match the effective Eddington time $ \eta\, t_E \approx 5 10^{-2}$ Gyr. 
In sum, the number of early QSs ought to track the galaxy formation, 
and {\it rise} together with the number of 
host galactic halos exceeding some $10^{10}\, M_{\odot}$. 

Such a direct connection of QSs with newly forming 
galaxies has been represented with models 
developing over the years from the first attemps  by Cavaliere \& Szalay 
(1986) and Efstathiou \& Rees (1988), to Haehnelt \& Rees (1993),  
Haiman \& Loeb (1998), Haehnelt, Natarajan \& Rees (1998), 
Cavaliere \& Vittorini (1998, henceforth CV98). Concurring evidence 
is being provided by observations such as those by   
Fontana et al. (1998), and in Djorgowski (1998). 

Here we focus on the dramatic 
{\it fall} of the bright QS population later than $z \approx 2.5$, where 
more features can be observed and the picture is more complex. 
The time scale involved is around $2$ Gyr, fast compared 
with the gentle demise of the galaxy formation envisaged 
by the hierarchical clustering. The behavior of the optical luminosity 
function (LF) is dominated by luminosity 
evolution (LE, see  Boyle, Shanks \& Peterson 1988; Hatziminaoglou, 
van Waerheke \& Mathez 1998), 
a trend also shared by the 
LFs in X-rays (Della Ceca et al. 1994, Boyle et al. 1994) and in the radio 
band (Dunlop \& Peacock 1990). 

Additional features appear in the optical data toward 
low $z$.  
First, a minor component of density evolution (DE) 
has been discerned at blue 
magnitudes $M_B \approx -23$ 
by Boyle et al. (1988), and confirmed by 
Maloney \& Petrosian (1999). 
Second, La Franca \& Cristiani (1997), and  
Goldschmidt  \& Miller (1998) found an excess of bright QSs over 
previous surveys; from these observations and those by 
K\"ohler et al. (1997) the LFs appear to be 
stretched out into a generally smoother shape on approaching low $z$. 
Finally, faint X-ray emissions and liner-type activity 
are increasingly found to be widespread among otherwise normal galaxies,  
see Ho, Filippenko \& Sargent (1997); Miyaji, Hasinger \& Schmidt (1998); 
Giommi, Fiore \& Perri (1998).

On the other hand, sharp imaging of a number of bright 
QSs at intermediate $z$ has revealed   
host galaxies which are not really newly forming, but rather are 
surrounded by a complex environment. Some still exhibit morphological 
marks of a recent tidal interaction,  
in spite of the low surface brightness  and short duration ($\sim 1/2$ Gyr) 
of such features; others are found in close 
 association with comparable or fainter companions; a fraction even appear to 
harbor secondary nuclei  
(see Hutchings \& Neff 1992; Hutchings, Crampton \& Johnson 1995; 
Rafanelli, Violato  \& Baruffolo 1995; 
Hasinger et al. 1997; Bahcall et al. 1997; 
Boyce et al. 1998; McLure et al. 1998; Ridgway et. al. 1999).
In addition, recent statistical evidence (Fisher et al. 1996;  
 Jaeger, Fricke \& Heidt 1998; Wold et al. 1999) 
points toward a poor group environment also for 
radio-quiet QSs, perhaps poorer than for the radio-loud objects 
(Yee \& Green 1987, Lehnert et al. 1999). 

But the issue is not settled. Other observers detect at most marginal 
evidence that the environments of the bright QS are richer than the field 
(see Smith, Boyle \& Maddox 1995; Teplitz, McLean \& Malkan 1999; 
Croom \& Shanks 1998), and find 
low incidence of peculiarities 
in galaxies active at the levels of the Seyfert nuclei 
(Malkan, Gorjian \& Tam 1998). 
Many point out the need to enlarge and complete the QS samples, and to 
compare them with proper controls. 

Such a complex data situation warrants exploring 
what predictions follow from pursuing 
the physical connection of bright QS activity with host interactions. 
Our guideline (taking up Cavaliere \& Padovani 1989, Small \& Blandford 1992, 
CV98) will be that later than $z \approx 2.5$ 
the accretion feast ought to cease turning the 
QS rise into a descent, due to a gradual {\it change} in their environment. 
In fact, toward lower $z$ 
major merging events (with the associated gas imports into the host) 
become rarer and rarer at galactic scales; 
the prevailing dynamical events are best described 
as {\it interactions} between developed galaxies, occurring mainly in the small
groups which then begin to virialize. 
These events do not necessarily increase the dark halo masses, yet they 
can {\it destabilize} the gas in the 
hosts and trigger limited accretion episodes; 
these draw from the gas reservoirs in the hosts, and 
concur with starbursts in progressively exhausting them. 
But as exhaustion is approached 
other, lesser gas supplies become relevant for lower level activity, 
such as those provided from accretion of satellite galaxies 
that occours even in low 
density environments or in the field. 

We intend to single out and address directly the 
astrophysical processes involved, rather than base upon 
 numerical and phenomenological 
treatments as in Kauffmann \& Haehnelt (1999). Specifically, 
we shall simplify the dynamical history of the DM, to focus on the consumption 
and depletion of the gas reservoirs in the host galaxies.
First, in \S 2 and \S 3 we will examine the outcomes of 
encounters of the host with sizeable companions, 
occurring mainly in poor {\it groups}. Then in \S 4 we examine accretion 
of satellite galaxies with their additional but small 
gas supplies, and consider also other processes that  
can occur in the {\it field}. Finally, in \S 5 we summarize and 
discuss our conclusions, stressing implications and predictions.

\section{Quasar activity triggered by encounters in a group} 

Our study will be based on the standard hierarchical development 
of cosmic structures 
past the era of galaxy formation; poor groups, in particular, 
are expected to start condensing 
at a considerable rate later than $z \approx  2.5$ in viable cosmologies. 
\footnote {
For analytical evaluations we will use the critical cosmology with 
$h = 0.5$ and tilted cold DM initial perturbations; 
our numerical calculations will also cover the flat, 
$\Omega_o = 0.3$ case with $h=0.7$ and 
standard cold DM perturbations (see Bunn \& White 1997). 
We recall that in the former cosmology the standard 
hierarchical clustering implies  
the virializing masses to attain density contrasts close to 180 over the field  
and to scale up like $M \sim 10^{15} \, (1+z)^{-6/(n+3)}\, M_{\odot}$, 
while the velocity dispersions scale like $V \propto 
M^{1/3}\, (1+ z)^{0.5}$; here $n \approx -2$ is 
the power spectral index of the perturbations at group scales. }


We hold that the approximate equality at $z\approx 2.5$ of the 
QS peak with the group onset constitutes  
not a coincidence but rather an {\it intrisinc} feature, telling of a connection of the QSs 
with the environment at scales larger than galactic. 
Our starting point is that QS hosts in early, poor but dense groups are 
bound to {\it interact} with their companions.
The role of these dynamical events for nuclear activity
is again to destabilize the gas in the host, specifically causing 
it to lose angular momentum.  
So they trigger inflows that rekindle the 
existing but dormant BHs into 
intermittent, dwindling episodes of {\it supply-limited} accretion and 
emission. 

That encounters occur preferentially in groups is shown 
by many simulations (see Governato, Tozzi \& Cavaliere 1996, 
Athanassoula 1998). This may be anticipated (see CV98)
on the basis of the average time  
$\tau_r \sim 1/ n_g \, \Sigma \, V \sim $ Gyr between encounters, 
basically because in groups the galaxy number density $n_g$ is high.   
On the other hand, in poor groups the velocity dispersion $V$ only 
modestly exceeds the internal galaxian velocities $v_g$, so that the 
cross section for effective interactions (see Binney \& Tremaine 1987) 
is close to the 
geometrical value $\Sigma \sim \pi (r_g + r_g')^2$ 
related to the radii $r_g$ of the host and $r_g'$ of the interaction partner. 
The time scale for inflows to develop 
is of order 
 $\tau \approx r_g/v_g \sim 10^{-1}$ Gyr, the galaxian dynamical time. 

That such encounters can lead to substantial gas funnelling toward the nucleus 
is shown by a number of high resolution simulations comprising 
both the N-body treatment of the DM component and a hydrodynamic treatment  
of the gas (see Mihos \& Hernquist 1996, Barnes \& Hernquist 1998). 
These simulations show that over a few $10^{-1}$ Gyrs following a grazing encounter 
up to $50 \%$ of the gas orbiting in a galaxy at distances of kpc 
is driven  into the central  $10^2$ pc, the 
resolution limit. 
The result follows from a number of dynamical steps, as 
discussed by Mihos (1999). First comes the gravitational action from the 
partner, which depends on its mass $M'$ and impact parameter $b$ and lasts for 
a time $b/V \sim r_g/v_G \approx \tau$. 
Then follows the response by the DM and the baryonic components 
developing in the host over times $\tau$; this is 
possibly amplified by internal metastability conditions to 
form true bars. Finally, it  
ensues the loss by the gas of a considerable amount of orbital energy and 
angular momentum $|\Delta j/j|$, which triggers the gas inflow.
An internal loss of $j$ (that is, $\Delta j < 0$) may be anticipated, since the 
asymmetries and inhomogeneities induced in the baryonic and in the 
DM components do not superpose or align; so the gravitational coupling 
will transfer $j$ from the former to the latter, more massive component.

But most accretion episodes will be milder, originating from a fly-by 
of the host by a smaller or a comparable companion at impact parameters 
$b \magcir r_g$. 
From eq. (1.1) it is seen that in gas-rich hosts still 
containing $ m \sim 5\,10^{10} \, M_{\odot}$ of gas,  
a fractional gas mass $\Delta m/m \sim$ a few $ \%$ 
sent to the central BH over $\Delta t \sim \tau \sim 10^{-1}$ Gyr is enough to 
yield outputs approaching $5\,10^{46}$ erg$s^{-1}$. 

The actual gas amount made available by an encounter 
will {\it limit} the maximal luminosity attained by the QS before fading out.
The controlling factor will be $|\Delta j/j|$, which closely bounds from above 
the gas fraction $f$ funneled to a small inward velocity (see Gunn 
1977). In fact, the 
gas funneled inward may end up not only in accretion onto the central BH,  
but also in a less constrained nuclear starburst 
(see Sanders \& Mirabel 1996), or it may be mostly dispersed. 
We shall assume that about 1/3 of the destabilized fraction 
$f$ constitutes the fraction  $\Delta m/m$ actually accreted. 

It is convenient to rewrite the output in eq. (1.1) as 
$$L \sim \eta\, c^2 \Delta m/\tau \propto  m(t) \, f ~, \eqno (2.1)$$
apart from the factor $ \eta\, c^2/ 3\,\tau$ 
fixed for a given host. 
This is to visualize and tell apart the long-term 
trend and the fluctuating component in the gas accretion. The factor $m(t)$ 
{\it drifts down} on the long scale set by the time $\tau_r\sim$ Gyr between 
encounters; over the shorter scales 
$\tau \sim 10^{-1}$ Gyr of a fly-by the factor $m(t)$ stays nearly constant,  
but the other factor $f  \mincir |\Delta j/j|$ 
is statistically {\it distributed} following the distributions 
of the dynamical parameters of the fly-by events. 

\subsection{The shape of the LFs from encounter statistics}

Here we show how the {\it distribution} $P(f)$ of the fractions $f$  
gives rise to the shape of the LF, that we indicate with $N(L)$. 

We expect that many encounters can destabilize gas fractions of order  
$f_b  \approx |\Delta j/j |\sim 5\%$; these will lead to accretion of 
$ (\Delta m/m)_b \sim 2 \, \%$,  and to outputs up to 
$L_b \sim 5\, 10^{46}$ erg$s^{-1}$ in a gas-rich host. 
We shall see that they produce 
$L \, N(L) \sim$ const in this luminosity range. 
Higher luminosities require larger destabilized fractions $f>f_b$; 
in turn, these require closer encounters with 
larger companions, and these events are fewer.  
When we compute the corresponding distribution $P(f)$, we  
expect it to drop 
steeply, producing a steep bright end of $L \, N(L)$.

To show this, we consider a group   
and inside it we examine 
the interaction of a gas-rich host galaxy 
(with 
radius $r_g$ and circular velocity $v_g$) with a group companion  of 
mass  $M'$ on its orbit with relative velocity $V$ and
impact parameter $b$. 
The gas mass $m$ in the host, at equilibrium  
on scales $r $ of some kpcs with the velocity $v \sim (GM_o/r)^{1/2}$ 
enforced by the host mass $M_o$ within $r$, 
has the specific angular momentum $j =  v r \approx G  M_o/v$. 
The maximal $|\Delta j|$ obtains when the gas rotates slowly and responds 
strongly to the gravitational force from the companion. 
In such conditions (see Appendix A) we evaluate 
the time-integrated, relative change 
$|\Delta j/j|\approx  j\, M'/ M_o\, V\, b$, 
which easily approaches 1.
Using this to evaluate $f $ in eq. (2.1), 
we see that $L \propto M'/b$ holds, on neglecting over the time scale $\tau$ 
the slow decrease of $m$. 

So $P(f)$ as defined above is computed on convolving 
the distributions of the dynamical parameters $M'$ and $b$, namely, 
$p_1 (M')$ and by $p_2 (b)$ combined into 
$$P(f) =  
 \int\int dM'\, db\; p_1 (M')\;   p_2(b)\; \delta (f - A\;  M'/b) 
~, \eqno(2.2)$$ 
where the factor $A  \equiv j /M_o V $ is fixed for a given host 
and over times of order $\tau$.   
The distribution $p_1 (M')$ is dispersed over a wide range; 
we adopt the Press \& Schechter (1974) form which includes the power-law 
factor $M'\;^{- 1.85}$, from $M' = 5\, 10^{10} \,M_{\odot}$ 
up to the mass of the host, often the 
brightest group member.
\footnote{Similar results obtain 
on starting from a Schechter LF for the galaxies, see Zucca et al. (1997), 
converted to mass distribution with the use of $L\propto M^{1\pm 1/3}$.}
The scaling of $P(f)$ in the upper range of $f$ 
depends only moderately on $p_2(b)$. The latter reflects the poorly known 
geometry of the galaxian orbits in the group; but  when the 
group is small, with membership of a few to several bright galaxies, 
the range of $b$ extends by a factor of hardly $10$,  
from the minimal value about $r_g$ (which sets 
encounters apart from prompt merging, see Governato el al. 1996) 
to the upper value $ R_G\, {\cal{N}}_G^{-1/3}$ 
(which defines the effective binary 
encounters in terms of the group radius $R_G$ and the membership 
${\cal{N}}_G$). 

We represent in fig. 1 the probability $P(f)$ that we compute in 
the Appendix A from eq. (2.2) in 
three cases that bracket the variance expected. One is the  
simplest case ${\cal{A}}$  obtained on using a peaked distribution $p_2(b)$; 
we find that the shape in the main body of $P(f)$ follows the shape of 
$p_1(M')$, 
to yield $P(f)= 0.85\,(f/f_b)^{-1.85}f_b^{-1}$ 
before going into a cutoff. 
The second case ${\cal{B}}$ is the other limit holding for a flat distribution 
$p_2(b)$, and yielding a shape still close to $P (f)\propto f^{-1.85}$ 
before the cutoff; this will constitute our reference frame. 
Case ${\cal{C}}$ obtains when the gas rotation is fast and 
the gravitational torque from the partner can be 
described as a truly tidal effect (see Appendix A). 
Then integrating over the 
fly-by time we evaluate $|\Delta j/j|\approx  j\, M'/ M_o\, V\, b \times r/b$; 
the shape of $P(f)$ turns out to be somewhat flattened in the extended 
lower range of $f$. 

\vspace{8.cm} 
\includegraphics{cv00f1.epsf}
\vspace*{1.5cm} 
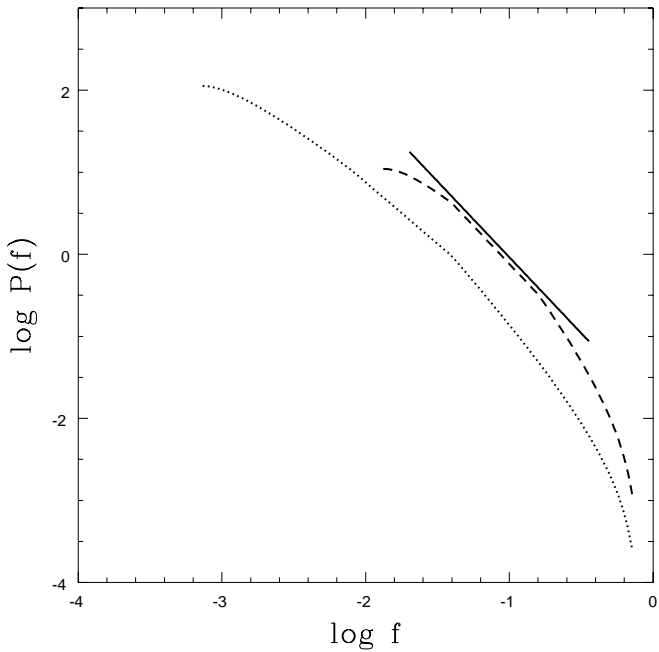
{\footnotesize FIG. 1 -- The probability distribution $P(f)$ 
that a fraction $f$ of gas is 
destabilized during an encounter, in the cases discussed in 
\S 2.1 of the text: ${\cal{A}}$ 
(solid line); 
${\cal{B}}$ (dashed line); ${\cal{C}}$ (dotted line). 
Press \& Schechter 1974 distribution for the 
partner masses from $M' = 5\, 10^{10}$ to $5\, 10^{12}\, M_{\odot}$. 
}
\vspace{\baselineskip}

To evaluate the corresponding shape of $ N(L)$, we consider that 
after times of order $b/V \sim 10^{-1}$ Gyr taken by a fly-by (see \S 2) 
the gas will increasingly 
flood the accretion disk in the nucleus on a galactic timescale $\tau$  
again of order $10^{-1}$ Gyr (see \S 2). The 
reactivated QSs will brighten up at the 
rate $\dot L = L/\tau$, up to the 
value of $L$ allowed by the accreted fraction $\Delta m/m \approx  f/3$; 
in the range $L<L_b$ corresponding to minimal $(\Delta m/m)_b$ 
their number is conserved, that is, $L\,N(L)\sim$ const applies. 
Then they fade off and drop out of the LF; 
the relative number of objects doing so in each range $dL$ 
around $L$ is given by the relative variation of $P$ as given by 
$$ d (N\, \dot L)/ N\, \dot L= dP/P~, \eqno (2.3)$$
where the right hand side is evaluated at $f \propto L$ 
(note the analogy  with the radiative transfer formalism). 
The result is 
$ L\, N \propto L^{-\beta}$ with $\beta\approx 2$ 
at luminosities brighter than $L_b$. 

In sum, past a break at $L_b \sim 5\, 10^{46}$ erg/s corresponding to 
$f=f_b\sim 5\%$, 
the QS luminosity function fed by 
encounters {\it steepens} from $N(L) \propto L^{-1}$ to 
$N(L) \propto L^{-1-\beta}$, 
and then goes into a cutoff. This evaluation will be checked with the 
systematic computation given in \S 3.

\subsection{Evolution from encounter rate}

The comoving LF evolves over the long time scales provided by time between 
encounters $\tau_r = 1/ n_g\, \Sigma\, V$.
The latter grows longer as $z$ decreases, basically because 
the density $n_g \propto (1+z)^3$  of galaxies in a group is proportional 
to the decreasing mass density in the background at virialization. 
This is offset only in part by the increase of the encounter 
velocities $ V\propto M_G^{1/3}\,(1+z)^{0.5}$ caused 
by the gravitational clustering, 
which approximately yields $V \propto (1+z)^{-3/2}$ 
in the critical universe where $M_G \propto (1+z)^{-6}$ holds.  
In the absence of strong galaxy evolution 
the cross section  remains close to $\Sigma \approx 4\pi r_g^2 \approx $const 
in groups or poor clusters. Then 
$\tau_r \propto (1+z)^{-3/2} \propto t$ 
closely obtains in groups. 

As a consequence, two components to the QS evolution may be  anticipated. 
A first, milder component is due to 
the encounters becoming 
{\it rarer} as richer, but less dense groups come of age. 
This clearly causes 
 in the QS population a mild DE proportional to $\tau_r^{-1} \propto 
(1+z)^{1.5}$ at late epoch $z\mincir 1$. 

The second and stronger component is in the form of LE; it occurs 
because on long time scales the luminosities decrease together 
with the available gas mass, following 
$L\propto m(t)$. In other words, 
the interactions also become {\it less} effective in 
feeding the BHs as the host gas reservoirs are  depleted by 
all previous discharges.

From $P(f)$ given by eq. 2.2, we compute 
$\langle f\rangle \, = \int df\, f\, P(f) \approx 15\%$ corresponding to 
an average $\Delta m/m \sim 5\%$ and to a minimal fraction  
$(\Delta m/m)_b \sim 2\%$ actually accreted. 
So the gas is {\it depleted} 
following
$$dm/dt  \approx - \langle f\rangle \, m / \tau_r(t)~, \eqno (2.4)$$ 
to yield $m(t) \propto (t/t_o)^{-\langle f \rangle\, t_o/\tau_{ro} }$, 
where $t_o \approx 13$ Gyr is the present epoch.
The value $\tau_{ro}$ of the time between encounters in local, virialized 
systems may be scaled from 
the classic census in the local field of bright galaxies 
with persisting signatures of interactions (Toomre 1977); 
this leads to a time around $2 \, 10^2$ Gyr between such encounters 
in the field. The value scales with the inverse of the galaxy 
density down to $\tau_{ro} \approx 1$ Gyr in virialized groups,  
where density contrasts  of about $2\, 10^2$ over the field 
are indicated by the hierarchical clustering. 
The resulting behavior is a  strong LE; in particular,  
in the critical universe the break luminosity $L_b\propto m(t)$ 
evolves as 
$$L_b \propto t^{-2} \propto (1+z)^{3}~.\eqno (2.5)$$ 
To sum up the contents of this Section, from direct evaluation 
of the rates and strength of the accretion episodes  
driven by encounters we have found the following 
features: the comoving number of the QSs  
evolves like $N \propto \tau_r^{-1} \propto (1+z)^{1.5}$ for $z<1$, the 
era of near completion for group formation, see \S3; 
the shape of the LFs has the form of a double power law broken at 
$L_b \sim 5\, 10^{46} [(1+z)/3.5]^3$; 
the bright end is given by
$L\, N(L) \propto (L/L_b)^{-\beta}$ with $\beta \approx 2$. Collecting 
these results in one expression, we expect the LF of QSs to follow 
$$N (L, z) \propto (1+z)^{1.5}\, L^{-1}\,\times \, (L/L_b)^{-\beta}~,  
\eqno (2.6)$$
with the last factor applying only beyond the break $L_b$. 
These basic behaviors of $N(L,z)$ will be 
checked and completed in the next Section.

\section{The full LF from population kinetics}

Here we present a systematic computation of $N(L,t)$, 
the differential and evolving LF in comoving form, 
produced when the BHs are reactivated by encounters to bright QS 
luminosities. We will base 
our computations upon the formalism of the so-called continuity equation 
(see Cavaliere et al. 1983).  

This pictures the statistics  of the 
QSs as analogous to a hydrodynamical flow along the  $L$-axis 
with sources and 
sinks. In fact, the equation that we are going to recall and use is the 
analog (whence the name) of the continuity equation 
for a 1-D fluid, with the density replaced by the comoving number of objects 
per unit luminosity $N(L, t)$, and the velocity replaced 
by $\dot L$. The equation stems from the general expression for 
the rate of change $\partial_t N$ in a given bin 
$dL$ around  $L$; the change is due to three causes. First, the objects are 
activated at the luminosity $L$ and with the rate given by a source function 
$S_+(L, t)$. Second, they brighten up at the 
rate $\dot L$ on the time scale $\tau$ discussed in \S 2, and 
so flow out of the bin $dL$ at the net rate $-\partial_L(\dot{L}N)$. 
Third, they may quench for lack of fuel 
after attaining the luminosity $L$, at the rate given by 
a sink function $S_-(L, t)$. 

The algebraic sum of the above contributions yields 
the net rate of change $\partial_t N$ in the form of the kinetic 
equation 
$$\partial_t N +  \partial_L(\dot{L}N) = S_+ - S_-~.  \eqno (3.1)$$
This is the ``continuity equation'' treated by Cavaliere et al. 1983.  
We now discuss in turn the three astrophysical inputs $\dot L,\, S_+$, and 
$S_-$ that substantiate the general formalism for the present scope. 

In the present context, the brightening $\dot L = L/\tau$ develops 
on the time scale $\tau \sim 10^{-1}$ Gyr (discussed in \S 2) 
taken by the gas to be directed on an inward course and to  
flood the accretion disk in the nucleus.  
 This implies  $L = L_i \, e^ {(t-t_i)/\tau}$, where  
$t_i$ is the epoch when reactivation and brightening begin from a 
faint luminosity $L_i$.
These ``characteristic lines'' $L(t)$ are related to the streamlines 
in our hydrodynamical
analogy, as shown in Appendix B; there we also show that they 
provide a simple way to find the relevant solution of eq. (3.1). 

The source function $S_+ (L,t)$ 
represents the rate of BH reactivations per unit $L$. 
As to its $L$-dependence, reactivation implies scarce if any accretion 
occurring previously,
so the initial luminosities $L_i$ are faint 
compared to the bright levels we focus on; hence to a first 
approximation we shall necglect the dispersion, and 
take $S_+(L)$ to be concentrated around a single value $L_i$ in the form of 
a $\delta$-function $\delta (L-L_i)$. As to the $t$-dependence, the 
reactivation rate is 
given by the (comoving) number density of hosts located in groups 
at the epoch $t$, 
divided by the mean time $\tau_r$ between reactivations.  
We focus on the member in a group 
that is so large and gas-rich as to fuel 
a sizeable activity upon encounters;  
often this will be single, based on the magnitude difference between the 
members of small groups observed by Ramella et al. (1999).  
So the number density of effective hosts will equal $N_G (t)$, the 
density of groups in the universe at the epoch $t$. 
 In sum, our source function is given by 
$$S_+(L,t)= N_G(t)\, \delta (L-L_i)/\tau_r(t)~. \eqno(3.2)$$  

Most rekindled sources attain $L\sim L_b \sim 5\, 10^{46}$ erg s$^{-1}$ 
as said in \S 2. 
So at {\it faint} luminosities $L < L_b$ no quenching occurs, and only 
the source term $S_+$ is present in eq. (3.1); then 
the solution of eq. (3.1) takes the simple form (derived in Appendix B, see
eq. B.6) 
$$N(L,t) = {\tau\, N_G(t) \over \tau_r(t)}  L^{-1} \, (L/L_i)^{-\gamma}~.  
\eqno(3.3)$$
Here we see the basic shape $N(L) \propto L^{-1}$ (corresponding to $L\, N  
\approx$ const, 
as expected from our preliminary discussion of \S2.1); this  
reflects the object population established 
by the brightening along the lines 
$L = L_i \, e^ {(t-t_i)/\tau}$. The modulating 
exponent $\gamma$ (evaluated in Appendix B) takes values close to 
$\gamma = \tau/t_G \approx  0.2$  
at epochs close to $t_G$  (i.e, at $z\approx 2.5$), when 
the group number rises briskly. This is because 
at such epochs the source function $S_+ \propto N_G(t)$ 
increases appreciably over the time scale $\tau$, so 
it tends to populate the faint bins close to $L_i$ faster 
than the objects can flow out 
to brighter bins, and $N(L)$ steepens somewhat. 
At $z\mincir 1$ instead, $N_G(t)$ is on its slow demise;  
so the exponent $\gamma$ vanishes or becomes slightly negative. 
At these late epochs 
the $t$-dependence of the LF is dominated by 
the decreasing factor $\tau^{-1}_r(t) \propto t^{-1}$  
which yields a weak and late DE; this is maximized in the 
critical universe where it follows $N \propto \tau_r^{-1} \propto (1+z)^{1.5}$. 
But we recall from eq. (2.6) that for $ z\mincir 2.5$ 
a stronger LE of the form $L_b\propto (1+z)^3$ is driven by the depletion 
of gas accreted or condensed into stars following each interaction. 

For $L > L_b$ also the sink function $S_-(L,t)$ intervenes 
at the right hand side of eq. (3.1); this represents the number of 
objects per unit time that brighten beyond $L_b$, but then 
quench and drop out of the LF at some larger $L$ due to fuel limitations. 

If these independently affect the QS population at a given $z$, then 
$S_- \propto N$ is to hold.
On the other hand, the drop out rate is related to the probability 
$P(f)\propto f^{-\beta}$ 
discussed in \S 2.1; 
clearly, a steeper slope $\beta$ 
will corresponds to faster drop out rates compared with the scale $\tau$ 
of the brightening.  
These considerations lead to  
$S_- =$ $N \, \beta/ \tau $; 
in other words, $\tau/\beta$ is the QS lifetime at high luminosities 
before fading out. 
Since $\beta =|d\, lnP/d\, ln f|$, and recalling that $\dot L =L/\tau$ holds, 
the relation also writes 
$$ S_-(L,t)=N(L,t)\,|\dot L\;  d\, lnP/L\; d\, ln f| ~.  \eqno(3.4)$$
This is just the same as given by eq. (2.3) under the form 
$d (N\, \dot L)/dL =N\, \dot L\, dP/dL\,P$. 

The solution for $L> L_b$ when both $S_+$ and $S_-$ apply is derived in 
Appendix B, eqs. (B.7) and (B.8), and  
yields the {\it bright} end behavior 
$$N(L,t) = {\, \tau\, N_G(t)\over \tau_r(t)\, L_b}  \, (L/L_i)^{-\gamma} \,
(L/L_b)^{-\beta-1} ~. 
\eqno(3.5)$$

The {\it full} shape of the LFs is computed in Appendix B using eq. (B.7).  
Specifically, if for $S_-$ we use the representation given by eq. (B.9) 
with $\beta=$const, then the solution is analytic and reads 
$$ N(L,t)={\tau\, N_G(t)\, (L_i/L_b)^{\gamma} \over \tau_r(t)\, L_b} 
{1\over (L/L_b)^{1+\gamma}+(L/L_b)^{1+\gamma+\beta}}~, \eqno(3.6)$$
which gratifyingly coincides with the empirical 
fitting formula used, e.g., by Boyle et al. 1988, 
Pei 1995, La Franca \& Cristiani 1997, 
Shanks at al. 2000.
Note that this goes over smoothly 
from the limiting form (3.3) at the faint end to the other limiting form 
(3.5) at the bright end.   
Note also that the natural normalization of the LF given by eq. (3.6)  
turns out to be a few $\%$ shining QSs per bright galaxy; this 
arises with no parameter adjustement from the factor 
$\tau/\tau_r\sim 10^{-1}$, and from the fraction up to 40\% of bright galaxies 
residing in groups with membership ${\cal{N}}_G \geq 3$ (Ramella et al. 1999) 
which yields $N_G \sim 0.4/3$ relative to the bright galaxies. 

\vspace*{2.8cm}
\vspace{15cm}
\includegraphics{cv00f2.epsf}
\vspace*{-.7cm}
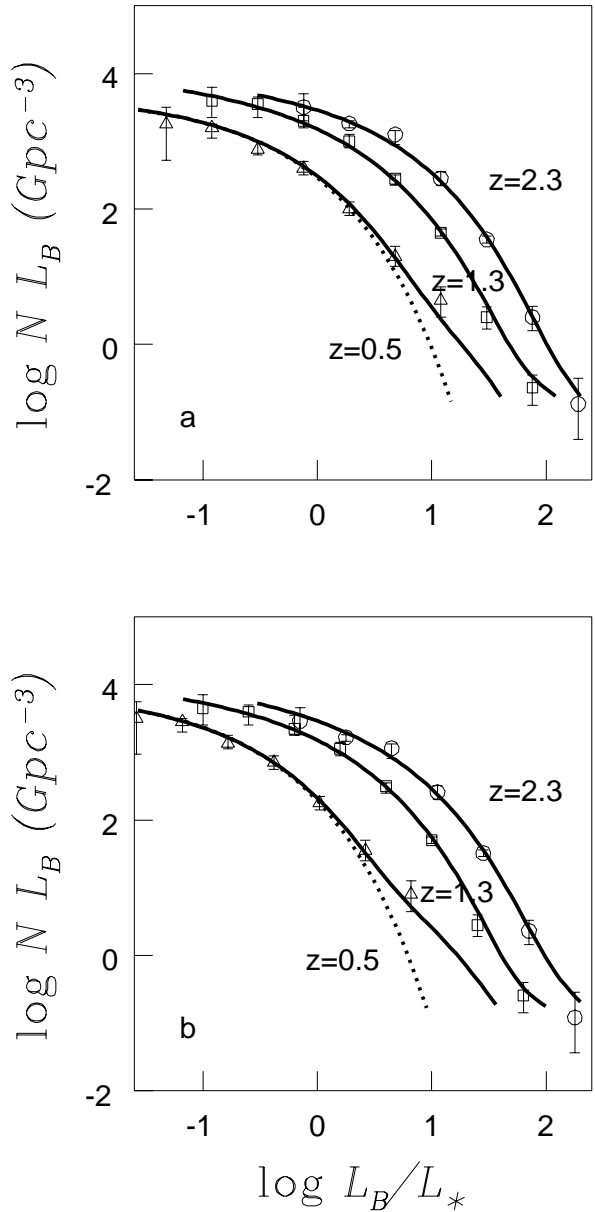
{\footnotesize FIG. 2 -- Optical LF from interactions of
the BH hosts in groups and in the field; the solid lines show the sum
(see \S 3 of the text). Efficiency
$\eta = 10^{-1}$, bolometric
correction for the blue band $\kappa_B = 10$, $L_*=10^{45}$erg/s.
a) Critical cosmology with $h=0.5$, tilted CDM perturbation spectrum.
{\it b)} Same for $\Omega_0=0.3,~\Omega_{\lambda}=0.7, h=0.65$,
CDM perturbations.
Data points from Boyle et al. 1988, and La Franca \& Cristiani 1997.
}
\vspace{\baselineskip}

In fig. 2 we represent the results of numerical computations 
from eq. (B.7) where we use the detailed numerical function 
$P(f)$ (with $f=3\,L\,\tau / \eta\,m\,c^2$) given in fig. 1, 
case ${\cal{B}}$.  
All computations are performed both 
in the critical universe with $\Omega =1$ (fig. 2a), and in 
the low density, flat cosmology with $\Omega_o =0.3, \, \Omega_\lambda = 
0.7$ (fig. 2b).  These lead to comparable 
results; actually, the latter case accords marginally better 
with the  data for QSs 
down to blue luminosities  $L_B \sim   10^{44}$ erg s$^{-1}$. 

\section{Fueling processes in the field}

In fig. 2 we have added to the bright end of the LFs two small contributions 
which arise even when the QS hosts are located in the field,  
due to the following processes. 

(i) BH fueling from 
encounters of their host with a substantial partner; these occur mainly in the 
large-scale structures which contain most galaxies and have 
contrasts of a few units over the general average 
as shown by redshift surveys  
(Ramella, private communication). Such galaxy densities are still  
lower by about $10^{-2}$ 
compared with those in virialized groups (see footnote $^{(1)}$), 
so the field encounters take place 
at the correspondingly lower rate $10^{-2}\, 
\tau_r^{-1}$. On the other hand, they  
involve more hosts by a factor about ${\cal{N}}_G/0.40 \approx 7$; 
this arises (see the discussion at the end of \S 3) because 
the most effective groups have membership ${\cal{N}}_G \magcir 3$ 
and comprise about $40\%$ of the bright galaxies. 
Thus the contribution from field encounters is around $7\, 10^{-2}$ relative 
to encounters in groups. 

(ii) A lesser contribution (around 5 \%, as discussed by CV98) is due to 
BHs flaring up to Eddington luminosities in the minority of galaxies  
still formed after $z \approx 2.5$ by the few 
major merging events still occurring. 

These two contributions are small but decrease slowly, because  
field galaxies undergo rare encounters and their gas 
is used mainly by the slow quiescent star formation 
(see Guiderdoni et al. 1998). 
So on approaching low $z$ such contributions are bound to emerge 
at the {\it bright} end, and to exceed 
there the rapidly decreasing contribution due to encounters in groups  
(singled out by the dotted lines in fig. 2). 

On the other hand, nuclear activities 
fainter than $L_B \sim  10^{44}$ erg s$^{-1}$ are less demanding 
in terms of accretion, 
and may feed on a number of different and superposing processes. 
To complete the above picture, we examine here 
the minor host-satellite interactions 
which become relevant when the gas reservoirs in the hosts approach exhaustion. 
From these events we anticipate a considerable addition to the {\it faint} 
end of the LFs; our expectations are outlined 
below, and are all checked and expanded in Appendix C. 

Since also the host galaxies located in the field can accrete 
their satellites, 
many more BHs can be activated by accretion related to such events. 
But lower activity levels $L < \eta_{-1} \,10^{45}$ erg/s 
will be so allowed. This is because of two concurring reasons: first, 
the satellites masses are small, in the range 
$M_s \sim 10^9 - 5\, 10^{10}\, M_{\odot}$ 
complementary to that discussed in \S 2.1; 
second, the usable gas supplies are provided mainly by 
the satellite cores, as we shall 
see below, and these contain meager gas amounts $m \mincir  10^{-3} \, M_s$. 
Very sub-Eddington $L$ will be produced in this way, 
since the central BHs in the hosts are already grown up. 
Another expected feature is that the distribution of the gas masses 
to be accreted from the satellites should be as steep as, or even steeper than  
the distribution of the faint galaxy masses. 
In conclusion, 
the associated LFs will be {\it steep}, and their 
normalization will be high due to 
the many field galaxies also activated in this mode, though at faint levels.  

On the other hand, 
the event rate will decrease sharply as satellites are accreted 
out of the initial retinue and their residual number per host 
galaxy is depleted. This will yield a {\it strong} DE (in fact, one which is 
$L$-dependent as we shall see), but will cause no LE since the satellite 
gas supplies will be accreted either all or nothing. 

Finally, as the accretion rate diminishes, the radiative efficiency 
itself may decrease below $\eta \sim 10^{-1}$; moreover, 
the optical emission 
is expected to be especially weak and often drowned into the galactic light, 
while the associated emissions in X-rays will be stronger and 
also more easily pinpointed. 
These features arise because regimes of ADAF or ADIOS kind (see the 
discussion by 
Blandford \& Begelman 1998 and references therein) are to set in 
at low accretion rates such as those fed only by satellites. 
While it is much debated which regime actually prevails, 
both will yield a low 
overall efficiency in using and/or converting into radiation 
the gas that reaches the 
accretion disk. In both cases 
the hard X-rays  should dominate over the 
soft X-rays, even more over the optical emissions, 
see Di Matteo et al. (1999). So at these low accretion rates the soft 
X-ray and the optical bands comprise 
a small fraction of the small bolometric luminosity actually radiated. 

Our detailed computations are given in Appendix C.
Here we stress that both the gas supply 
carried by a satellite, and the gravitational perturbation affecting the host, 
will be effective for accretion only when 
a satellite or its residual core sinks deep into the host body and 
delivers there the gas to be accreted. 
Such sinking may be triggered by encounters, but more often 
will begin as orbital decay caused by 
dynamical friction; the formal time scale is given by 
$\tau_s = 3\, (M_s/5\, 10^{10}\, M_{\odot})$
Gyr, see Binney \& Tremaine (1987). 

However, the effective time scale actually depends on 
the current satellite mass $M_s(t)$; it has been recently 
stressed how thoroughly the sinking satellites will be 
peeled off along their way by tidal disruption and stripping, 
especially those which are initially larger,  
see Walker, Mihos \& Hernquist (1996) and Colpi et al. (1999).  
The full process is very delicate to calculate and even to 
simulate numerically,  but it basically leads to 
a {\it weak} dependence on the initial 
$M_s$. 
In view of these uncertainties, our 
scope here will be limited to explore trends; to this 
specific purpose we adopt the phenomenological scaling  
$$\tau_s = 3\, (M_s/5\, 10^{10}\, M_{\odot})^{-\xi}\; Gyr \eqno (4.1)$$
with the exponent $\xi < $ $1$ designed to describe a 
weak dependence on the initial 
$M_s$. 

For consistency, we have also to consider that 
the gas effectively sunk into the host body 
is provided only by the fraction residing in the satellite cores; this 
is not only small but also will depend weakly on the initial total 
mass $M_s$. 
Again phenomenologically, we will use the exponent $\theta < 1$ 
to describe this dependence in the form 
$$ m = 5\, 10^{7}\, M_{\odot} (M_s/5\, 10^{10}\, M_{\odot})^{\theta}\; Gyr 
~. \eqno (4.2)$$

The full LFs so generated are computed in Appendix C, and given in eq. (C.5), 
using the appropriate form of the kinetic equation (3.1).
To illustrate the generic behaviors we give 
the result corresponding to the specific values $\xi= \theta= 0.7$, which 
yield 
$$N(L,t) \propto L^{-2.2}\; 
e^{- L\, t/3\, L_s }
~. \eqno (4.3)$$
Here $L_s \sim 10^{45}$ erg s$^{-1}$ is  the maximal bolometric 
luminosity from the gas mass $m = 5\, 10^7 M_{\odot}$ 
accreted from the core of a large satellite which spirals 
into the host on the scale $\tau_s \approx 3$ Gyr; the time 
$t$ is also given in Gyrs.
 
\vspace{8cm} 
\includegraphics{cv00f3.epsf}
\vspace*{1.5cm} 
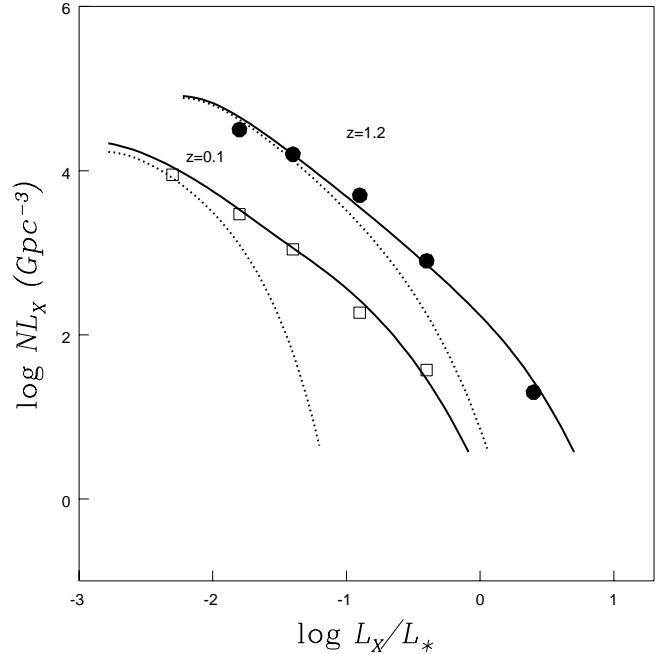
{\footnotesize FIG. 1 -- The contribution (dotted lines) to the X-ray LFs in the keV range
by accretion of 
satellites with masses below $5\, 10^{10} \,M_{\odot}$; we have taken 
$\eta/\kappa_X = 10^{-3}$ as discussed in \S 4 of the text. 
This contribution adds as shown by the solid lines to the 
one due to encounters in groups, represented 
in fig. 2a. Data points from Miyaji et al. 1998. 
}
\vspace{\baselineskip}

In fig. 3 we represent specifically the 
soft X-ray LFs at low $z$ resulting in the critical universe 
from cannibalism of satellites. 
We have adopted 
$\eta / \kappa_X$ $\approx  10^{-3}$, considering (as commented above) the 
small efficiency $\eta < 10^{-1}$ 
and the large bolometric 
correction $\kappa_X > 10$ expected for soft X-ray emission 
 at low accretion rates.  
We represent also how these LFs add to the 
contribution from group encounters that has been computed in \S 3 and 
represented in fig. 2a.

In sum, the contributions from satellite cannibalism 
to the LFs tend to pile up toward the faint end, producing steep LFs 
which steepen yet to a cutoff. Such contributions will 
evolve following a strong, $L$-dependent DE.

\section{Conclusions and discussion} 
We have shown that a steep QS fall for $z< 3$ is to result 
from the effects of the hierarchical {\it growth} of structures around 
accreting BHs, amplified by the exhaustion of the gas available for the 
accretion. 

The hierarchical clustering envisages  the era of 
galaxy formation to go over to an era of group assemblage for $z \mincir 2.5$. 
In a group, the settled 
host galaxies keep on evolving for a while, no longer by 
major merging events but rather via {\it encounters} which 
occur at rates $\tau_r^{-1} \sim n_g\, \Sigma\, V \sim$ 
Gyr$^{-1}$. 

These are milder events, yet able to destabilize the host gas, specifically by 
causing considerable {\it loss} of angular momentum as tackled in \S 2.   
We have shown how they trigger intermittent and {\it supply-limited} 
accretion episodes onto the 
central BH, yielding blue luminosities $L_B \magcir  10^{45}$ erg/s  
for some $10^{-1}$ Gyr in large, gas-rich galaxies. 
But we have additionally 
shown in \S 2.2 how such substantial gas discharges, which also 
fuel simultaneous starbursts activity, {\it deplete} the 
host reservoirs and cause the gaseous mass to 
decrease substantially following $m(t) \propto t^{-2}$ on average. 

Thus the  encounters not only turn rarer ($\tau_r \propto t$) 
 as the clustering proceeds toward groups richer but less dense, 
and so produce a later, {\it weak} DE; but they also 
become less and less effective in terms of triggered accretion, and 
this causes {\it strong} LE with 
the luminosities scaling down close to $L_b \propto (1+z)^3$.
The result we predict for bright QSs in the range  $z \mincir 2.5$ 
is outlined in eq. (2.6),  
is derived and shown in more details in \S 3 to yield eq. (3.6), and is 
represented in full by fig. 2.
This agrees not only with the observations by Boyle et al. 1988, but also with 
the recent re-analysis of variuos samples by Maloney \& Petrosian 1999, 
and with the greatly extended samples being provided at $M_B < -23$ 
by Shanks at al. 2000.

As to the shape of the LFs, we have shown in \S 2.1 
that the statistics of the encounters 
naturally yields a double power-law {\it broken}  from 
about $L^{-1.2}$ toward the faint end, to $L^{-3.2}$ and {\it steeper} 
at the bright end, in accord with the above data. But 
toward low $z$ a {\it smoother} overall shape is produced 
by the changing tilt $\gamma$ given by eqs. (3.6); this  
concurs with the enhancement of the bright end at $L_B \magcir 10^{46}$ erg/s 
provided by the additional feeding processes in the field
discussed in \S 4 and shown in fig. 2. 
Our result accords with the observations 
by La Franca \& Cristiani (1997) and by Goldschmidt \& Miller (1998), 
which are confirmed by a recent survey being analyzed by Cristiani 
and collaborators (see Grazian et al. 2000). 

The above values and behaviors are given for the critical universe, see 
fig.2a; but 
fig. 2b shows that similar or better results obtain in a flat, low density 
cosmology. 

From a different path, we agree  with Cattaneo, Haehnelt 
\& Rees 1999 and Kauffmann \& Haehnelt 1999 that strong QS 
evolution down to very low $z$ implies destabilization of 
substantial fractions of the gas stockpiled in the hosts. 
Only a minor part of such gas ends up into accretion episodes; 
the major part is likely to be heated and/or ejected, or 
used up to produce -- over and above the 
quiescent star formation -- star bursts specifically 
in the bright host galaxies, 
mostly ellipticals. But we see reasons why the LE evolution 
component may be expected to slow down on approaching $z\mincir 0.5$. 
For example, when interactions cause outright galaxy aggregations 
the host mass $M_g$ grows appreciably while its 
internal density $\rho_g$ decreases; then it is easily checked that 
the scaling of $\tau_r$ includes the factors 
$M_g^{1/3}\, \rho_g^{2/3}$, 
giving rise to $\tau_r \propto t ^{1+ \epsilon} $ with $\epsilon < 1$. 
Correspondingly, from eq. (2.4) the late evolution $L\propto m(z)$ 
brakes into a softening exponential  
at low $L$ and $z$.

On the other hand, when it comes to faint AGNs at $z \mincir 0.5$  
the situation becomes more complex, since the underlying BHs 
may be fed at low levels by various processes which superpose. 
We have explored in \S 4  and in fig. 3 
one such process, 
namely, the {\it accretion} of satellites with the associated 
gas supplies; this contributes 
many more but weaker sources, in the range of the Seyfert galaxies 
and below. These sources undergo a strong 
$L$-dependent DE, similar to that observed in X-rays by Miyaji, Hasinger \& Schmidt 
(1998);  many, in fact, will be more easily pinpointed in X-rays  
at $L_X \sim 10^{43}$ erg/s or fainter.  

Other feeding processes are conceivably 
provided by the self-destruction of giant nuclear star clusters 
 (see Norman \& Scoville 1988), and by internal instabilities and bar 
formation in disk galaxies (see 
Heller \& Shlosman 1994, Merritt 1998, Sellwood \& Moore 1998). 
It will be interesting to see what specific shape and evolution are 
predicted for the LFs of AGNs fed in this latter fashion. 

Here we focussed on {\it dynamic} losses of the gas angular momentum $j$
as primary triggers of the inflow eventually leading to 
BH accretion; the gas 
may undergo other adventures before actually fueling the 
central BH, see e.g. Siemiginowska \& Elvis (1997). But the 
straight prediction from our study
 substantiating the outline by Cavaliere \& Padovani (1989), 
is that the emissions from the average QS or AGN should go 
progressively {\it sub-Eddington} later than $z \approx  2.5$, 
even at constant 
$\eta \sim 10^{-1}$; this accords with the data trend taking shape 
from various observational lines, see Sun \& Malkan (1989), 
Haiman \& Menou (1998), Salucci et al. (1998), Wandel (1998). 

Our specific prediction for the average Eddington ratio is 
$$L/L_E \approx  [(1+z)/3.5]^{3} \, M_{BH}(2.5)/ M_{BH}(z)~, \eqno(5.1)$$
which goes down to about $10^{-2}$ by $z\approx 0.1$.
 This is based on the intrinsically 
{\it depleted} accretion onto already existing, massive BHs which yields 
$L \propto (1+z)^3$ as discussed above; the other factor is the related  
mass growth by the ratio $M_{BH}(z)/M_{BH}(2.5)$, see CV98. 

In fact, as outlined there and expanded elsewhere, by all 
episodes of rekindled accretion in the range $z \mincir   2.5$ 
the central BHs now dormant in most galaxies 
grow by an average factor $M_{BH}(0)/M_{BH}(2.5) \approx 5$. 
This will be lower for hosts which are disk dominated at radii of a few kpcs, 
based on the values of $\Delta j/j$ indicated in \S 2.1 by 
case ${\cal{C}}$ compared with case 
${\cal{B}}$. Otherwise, the factor will be limited by the bound 
$M_{BH} \mincir 10^{-2}\, M_{bulge} $ set when a large central condensation 
enforces axial symmetry despite the dynamical perturbations. 
Correspondingly, the local 
density of baryons collapsed into BHs amounts to 
$\rho_{BH} \approx 3\,10^{15}\, h^3 \, M_{\odot}$ Gpc$^{-3}$. 
These findings agree with the observational estimates of 
the central BH masses in many local galaxies given by Richstone et al. 1998,  
and with the upper bounds set by Magorrian et al. 1998.

Toward a comprehensive picture, we point out that the 
losses of $j$ during galaxy-galaxy interactions  
as here analyzed go back to dynamically generated {\it asymmetries} of 
the gravitational potential. Compared with the major merging events during 
the build up of the spheroids, these 
do not differ in kind, but only as for the lesser strength and 
as for their falling rate and decreasing gas amounts involved. 

So we conclude that 
most of the bright QS evolution can be understood in terms of 
{\it one} engine (the accreting BH) and {\it one} trigger 
(dynamical destabilization of the gas),  
which however operate under {\it two} adjoining regimes geared to the 
development of cosmic structures. Such regimes are related to 
the early, violent merging events when 
the gas-rich spheroids form; and to the later, 
dwindling interactions of host galaxies in groups or even in the field, 
with the slow demise of galaxy formation 
magnified by the fast depletion of the gas supplies.

\bigskip

We thank for helpful exchanges B. Guiderdoni, 
 B. Liberti, P. Rafanelli and M. Ramella. 
Early stimulating discussions with G. Hasinger and C. Norman 
are gratefully ackowledged. 
Thanks are due to our referee for stimulating us to clarify our 
presentation throughout the MS.
Work supported by partial grants from ASI and MURST. 

\section*{Appendix A}

As said in \S 2.1 of the main text,  
we consider in a group of radius $R_G$ and membership ${\cal{N}}_G \geq 3$ a 
gas-rich host 
galaxy with radius $r_g$ and circular velocity $v_g$, and a companion of 
mass  $M'$ flying by with 
impact parameter $b$ and relative velocity $V$. 
The gas in the host 
has specific angular momentum $j =  v r \approx G  M_o/v$ 
if it is at equilibrium  
on scales $r\sim $ kpc  with the velocity $v \approx (GM_o/r)^{1/2}$ 
under the gravity provided by the host mass $M_o$ within $r$. 

Let us consider a ring of gas.
When the rotation is slow with $v/V < r/b$ 
the gravitational torque exerted by $M'$ on the unit mass 
may be evaluated from 
the angular derivative of the interaction energy 
$$ T = {\partial E_p \over \partial\phi} \approx
  {2\, G\; M^{\prime} \, r \over \pi \, b^2}~. \eqno(A.1)$$ 
The overall variation $\Delta j$ of the specific angular momentum 
will result from the 
torque time-integrated along the partner orbit:  
$$|\Delta j| =\int dt\, T \approx {G\, M^{\prime} 
\, r \over V\, b}~, \eqno(A.2)$$ 
to read 
$$ |{\Delta j \over  j }| \approx 
{M^{\prime} \,j  \over M_0\,V\,b}~.\eqno(A.3)$$ 
Approximately 1/2 of the gas is braked and 
flows toward the nucleus, while the rest flows outward. 

With accretion so triggered, the probability $P(f)$ of destabilizing 
a fraction  $f= A\,  M'/b$ before fading out is given by eq. (2.2) 
in the main text. 
In the limiting case (denoted by ${\cal{A}}$ in the text) of a peaked 
distribution $p_2(b)$, 
the integrations are easy to perform analytically, to find 
$P(f) = 0.85 \, (f/f_b)^{-1.85} f_b^{-1}$, having assumed 
the Press \& Schechter shape 
$p_1(M') \propto M'^{-1.85}$. This result is represented in fig. 1 by the 
solid line.

The opposite limit, case ${\cal{B}}$, obtains on considering a flat 
distribution $p_2(b)$ in the range 
$r_g \, -\, R_G \, {\cal{N}}_G^{-1/3}$. 
Then the numerical integration of eq. (2.2)  yields the dashed 
line in fig. 1. This accords with the analytic evaluation 
which provides the shape 
$P(f) \propto (f/f_b)^{-1.85}$ between the limits 
$A\, M'_{min}/r_g<f< A\, M'_{max}\;R_G \,{\cal{N}}_G^{-1/3}$, before 
the final cutoff. We have used $M'_{min} = 5\, 10^{10}\, M_{\odot}$
and $M'_{max} = 5\, 10^{12} \, M_{\odot}$, as said in \S 2.1 of the main 
text.

Case ${\cal{C}}$
is one of relatively fast rotation with $v/V > r/b$. Then the net torque may be 
computed from differentiating eq. (A.1) across the diameter. This 
brings in the tidal factor  $r/b$, so that  
$$ |{\Delta j \over  j }| \approx 
{M^{\prime}\, j \, r\over M_0\,V\,b^2}~\eqno(A.4)$$ 
now holds. The luminosities so produced are given by 
$L = \eta \, c^2 m(t)$ $\, M^{\prime}\, j \, r/ M_0\,V\,b^2\, \tau$. Note  
that $j\, r/M_o \, \tau $ is nearly independent of the host 
parameters; this is seen on recalling that $v^2 \approx GM_o/r$, and using 
the approximation 
$\tau \approx r/v$. The main dependences of $L$ are on $m$, which however 
stays nearly  constant over 
the time scale of a fly-by, and on the interaction parameters which are 
statistically dispersed as here computed. 

For the purpose of the 
convolution (2.2) we now have $f =  A\,r\; M'/b^2$, and as before we have to 
integrate over the distributions of $M'$ and $b$. 
Considering again a flat distribution for $p_2(b)$ we numerically obtain 
the dotted line in fig. 1. The analytic evaluation 
yields again a shape 
close to $P(f) \propto f^{-1.85}$, but now in the range 
$A\,r\,M'_{min}/r_g^2<f< A\, r\, M'_{max}/R_G^2\, {\cal{N}}_G^{-2/3}$; 
this is followed by a cutoff, and is preceded by a somewhat 
flatter section toward the extended lower end.

\section*{Appendix B}

Here we give the analytic solution of the kinetic equation eq. (3.1), 
namely of  
$$\partial_t N +  \partial_L(\dot{L}N) = S_+ - S_-~, \eqno (B.1) $$
where specifically $\dot L = L/\tau$ applies as explained in \S 3 of 
the main text. 

The simplest presentation is based on rewriting the $2^{nd}$ term on the 
left hand side in the form 
$\partial_L(\dot{L}N) = \dot{L}\;\partial_LN\;+\;N\;\partial_L\dot{L}$; so
we may look at the equivalent equation  
$$\partial_t N + \dot{L}\;\partial_LN = - N\;\partial_L\dot{L} + S_+ - S_-~. 
\eqno (B.2)$$
The advantage is to visualize that the two terms on the left hand side make up 
the differential operator $\partial_t\;+\;\dot{L}\;\partial_L$; this 
is the total time derivative along the lines (known as ``characteristics'',   
see Shu 1992) defined by 
the solutions $L=L_i\,e^{(t-t_i)/\tau}$ of the equation 
$\dot{L} = L/\tau$, where $t_i$ is the epoch when reactivation and 
brightening begin from the luminosity $L_i$.
Such a derivative, that we denote with a dot, corresponds 
just to the Lagrangian derivative along the streamlines in the hydrodynamical
analogy presented in \S 3.

Thus along the characteristics 
we can reduce our partial differential equation eq. (B.1) to 
the ordinary differential equation  
$$\dot{N}\;=-N{d\dot{L} \over dL}\;+\;S_+\;-\;S_- ~. \eqno (B.3) $$
This is $1^{st}$-order, linear and in canonical form, with solutions  
given in textbooks, see Dwight 1961. 

We are interested in the initial condition $N(L,t)=0$ for $t\rightarrow 0$; 
that is, very few QSs were activated by encounters in groups at early 
epochs corresponding to $z > 2.5$ when few groups existed.
We first write the solution in the {\it faint} 
range $L< L_b$  where $S_- = 0$ holds, as explained in \S 3; there we 
have  
$$ N(L,t)= \int_{0}^t \, dt'\,  S_+(L',t')\, e^{-\int_{t'}^t dt''\,
d\dot{L}/dL}~. \eqno (B.4)$$
Actually this expression simplifies a great deal because of two 
circumstances: first, 
$dt = dL/\dot L$ holds on a streamline, so that 
we may equivalently integrate over $dL$; 
second, the expression of $S_+$ 
eq. (3.2)  is proportional to  $\delta (L-L_i)$. These two 
features together make the integrations straightforward, and the result is 
$$ N(L,t)=  {N_G(t_i) \over  \tau_r(t_i) \, \dot L} ~. 
\eqno (B.5)$$

To make explicit the contents of this solution 
we substitute $\dot L = L/\tau$, and express 
$t_i$ in terms of $t$ using $t - t_i = \tau\,ln\,L/L_i$ along the 
characteristics. 
The difference $t-t_i$ matters close to the epoch of group formation 
$t_G$ (that is, $z \approx 2.5$) when the group number density rises rapidly,  
following closely $N_G(t) = N_{G} (t_i)\, e^{(t-ti)/t_G}$; 
so $N_{G} (t_i) = 
N_G (t)(L/L_i)^{-\tau/t_G}$ may be substituted in eq. (B.5), while 
$\tau_r(t_i) \approx \tau_r(t)$ holds. 
Thus the explicit solution reads 
$$N(L,t) = {\tau \, N_G(t)\over \tau_r(t)}  L^{-1} \, (L/L_i)^{-\gamma}~,   
\eqno(B.6)$$
where $\gamma =\tau/t_G$, This is the eq. (3.3) of the main text.
We end the study of the range $L< L_b$ by noting that 
the integrand $- d\dot{L}/dL$ in  the exponential of eq. (B.4) 
is just the  coefficient of the term proportional to $N$ in eq. (B.3). 

When we consider the {\it bright} range $L > L_b$, we see from eq. (3.4) 
that also the sink function 
$S_- = - N\; \dot{L}\;d\, lnP/L\; d\,lnL$ intervenes. Now 
the complete coefficient of $N$ reads 
$-d\dot{L}/dL + \dot{L}\;d\, lnP/L\; d\, ln L$, and this will appear  
in the exponential of the solution which now writes 
$$ N(L,t)= \int_{0}^t \, dt'\,  S_+(L',t')\, e^{-\int_{t'}^t dt''\,
(d\dot{L}/dL -\dot{L}\;d\, lnP/L\; d\, ln L) }~. \eqno (B.7)$$
A similar calculation applies as that described for eq. (B.4); the term 
added to the exponential will give rise to a factor multiplying the 
faint end form given in eq. (B. 5). 
Using the expression $P(L)\propto (L/L_b)^{-\beta}$ 
($\beta \approx 2$) found in \S 2, 
the result reads 
$$N(L,t) = {\tau \, N_G(t_i) \over \tau_r(t_i)\, L} \, ({L\over L_b})^{-\beta} ~. 
\eqno (B.8)$$ 
Recalling that 
$t_i$ may be expressed in terms of $t$ using $t_i=t-\tau\,ln\,L/L_i$ 
on the characteristics, we obtain eq. (3.5) of the main text.

To compute the full LFs, 
we use  for the coefficient $S_- (L,t)/N$ (which in fact constitutes 
the integrand in the second exponential of eq. B.7) a smooth 
transition across the break $L_b$. This start from 
zero at $L \ll L_b$ where $S_-$ vanishes (see \S 3 and above), 
and for $L > L_b$ 
has to go to the full value 
given by eq. (3.4). 
A convenient representation is provided by 
$$S_- / N =  \beta\, (L/L_b)^{\beta}/\tau \,[1+ (L/L_b)^{\beta}]
 ~. \eqno(B.9)$$ 
This may be used to integrate eq. (B.7); 
if we keep $\beta=$const, we obtain the  
analytic solution given by eq. (3.6), which has the form of
a widely used fitting formula for the LFs (see text at the end of \S 3). 

Alternatively, we may integrate numerically eq. (B.7) using  the numerical form 
of $P(f)$ given in fig. 1, to compute the LFs represented in fig. 2 of 
the text.  
\section*{Appendix C}

Here we compute in detail the contribution to 
the LFs from accretion of satellite galaxies, following up the discussion 
in \S 4.

We consider satellites of dynamical masses in the 
range $M_s \sim 10^9 - 5\, 10^{10}\, M_{\odot}$, 
with the Press \& Schechter (1974) distribution 
$N_s(M_s)\propto M_s^{-1.85}$. We assume a satellite to contain 
a usable gas mass $m \sim  10^{-3} \, M_s$, since only 
the gas residing in the 
core will be ultimately effective for BH feeding. 

We use the same basic formalism of the continuity equation 
for $N(L,t)$ as given 
in \S 3 of the main text by  eq. (3.1),  
 with $\dot L = L/\tau$. 
But here the form of source $S_+$ and of the sink function $S_-$ 
will require a number of changes. 
They go back to the following facts: here {\it all} 
dormant BHs in bright galaxies, whether in groups or in the field, may be 
activated; in addition,  
$\Delta m \sim m$ holds (the gas in the satellite cores 
is accreted {\it all or nothing}); finally, accreting these meager gas supplies 
yields relatively {\it low} bolometric outputs $L\sim  \eta\, mc^2/\tau$ 
bounded by the value $L_s \sim  \eta_{-1} \, 10^{45}$ erg s$^{-1}$ 
for a large satellite.                             

The source function $S_+$ is now proportional to the number density  
$N_{BH}$ of all BHs formed by $z =2.5$ which amount to 
about 1 per bright galaxy 
(see Haehnelt, Natarajan \& Rees 1998); but it is also proportional to 
the diminishing satellite number per host galaxy ${\cal{N}}_s(t)$.
These factors are divided by the time $\tau_s$ taken to completely accrete 
a satellite, see eq. (4.1). 
The result reads
$$S_+= N_{BH}\, {\cal{N}}_s(t)\, \delta (L-L_i)/\tau_s ~.\eqno(C.1)$$ 
In turn, the satellite number 
${\cal{N}}_s(t)$ is governed by  
$ d{\cal{N}}_s/dt = - {\cal{N}}_s(t)/\tau_s$, 
assuming the satellites are independently 
accreted. This yields the long term decrease 
$${\cal{N}}_s(t) = {\cal{N}}_{si} \; e^{- t/\tau_s} ~ \eqno(C.2)$$ 
from the initial retinue ${\cal{N}}_{si}$ that we conservatively take  
at ${\cal{N}}_{si} \sim 10$.

The form of the sink function $S_-$ again follows 
eq. (3.4), namely 
$$S_-(L,t)=N(L,t)\,|\dot L\;  d\, lnP/dL| ~. \eqno(C.3)$$
But now $P(L)\,dL=N(m)\,dm$ is given 
straighforwardly in terms of 
the distribution of the satellite gas masses, 
since $ L\propto \Delta m \approx m$ 
hold. As explained in \S 4 of the main text, 
we have to consider that the effective gas supply is provided 
by the satellite cores. Then  
the effective gas mass will depend only weakly on the total satellite mass, 
following  $m \propto 
M_s^{\theta}$ (with $\theta < 1$) as indicated 
by eq. (4.2) and related comments. 
So $N(m)\, dm \propto M_s^{-1.85}\, dM_s$ holds, and $P(L) \propto 
(L/L_b)^{-1 -0.85/\theta}$  obtains for bolometric luminosities exceeding 
$L_b \sim 10^{44}$ erg s$^{-1}$ corresponding to the lower end of the 
satellite dynamical masses.  

This we use in the solution of the kinetic eq. 
(3.1), already given in general form by eq. (B.7). 
Evaluating the integrals as described there, we find 
$$ N(L,t) 
= {\tau \, N_{BH}\, {\cal{N}}_s(t)
 \over \tau_s\, }\; (L/L_b)^{-2 -0.85/\theta}  ~. 
\eqno (C.4)$$

But following eq. (4.1) of the main text and related comments, 
we have to consider that $\tau_s \propto M_s^{-\xi}$ (with $\xi < 1$) 
holds together with 
the dependence $L\propto m \propto M_s^{\theta}$ used above; it follows that 
$\tau_s \propto L^{-\xi/\theta}$. When we make explicit this dependence and 
the decrease of ${\cal{N}}(t)$ given by eq. (C.2) 
the result reads 
$$N(L,t) =  {\tau \, N_{BH}\, {\cal{N}}_{si} \over 3} 
(L_b/L_s)^{1 + 0.85 /\theta} \; \times$$
$$(L/L_s)^{-2 +(\xi  -0.85) /\theta}\; 
e^{- (L/L_s)^{\xi/\theta}\,  t/3}
~ .\eqno (C.5)$$
Here all times are given in Gyrs; $L_s \sim 10^{45}$ erg s$^{-1}$ 
is the maximal bolometric luminosity activated by the accretion of 
gas mass $m = 5\, 10^7 M_{\odot}$ from the core of a large satellite of mass 
$M_s= 5\, 10^{10}\, M_{\odot}$ spiralling down on the time scale $\tau_s = 
3$ Gyr (see the coefficients in the eqs. 4.1 and 4.2). 

Note that 
the QSs or AGNs fueled 
by satellites outnunber those activated by encounters in groups by 
the  factor  $N_{BH}\, {\cal{N}}(t) \, \tau_r/ N_G\, \tau_s \sim 5$; this 
specifies the high normalization anticipated in \S 4. 
Eq. (4.3) in the main text obtains for the illustrative values 
$\xi=\theta=0.7$.

\end{document}